\documentclass[aps,prl,twocolumn,superscriptaddress,floatfix]{revtex4-1}
\usepackage{color}
\usepackage{array}
\usepackage{verbatim}
\usepackage{multirow}
\usepackage{amsmath}
\usepackage{amssymb}
\usepackage{dsfont}
\usepackage{graphicx}
\usepackage{threeparttable}
\usepackage{esint}
\usepackage[unicode=true,
 bookmarks=true,bookmarksnumbered=true,bookmarksopen=false,
 breaklinks=true,pdfborder={0 0 0},backref=false,colorlinks=true,
urlcolor=blue,linkcolor=blue,citecolor=blue] {hyperref}%
\usepackage{cleveref}
\usepackage{float}
\usepackage{longtable}
\usepackage{diagbox}
\usepackage{tabularx}

\begin{document}
\title{Classification of second harmonic generation effect in magnetically ordered materials}
\author{Rui-Chun Xiao}  \email{xiaoruichun@ahu.edu.cn}
\affiliation{Institute of Physical Science and Information Technology, Anhui University, Hefei 230601, China}

\author{Ding-Fu Shao}
\affiliation{Key Laboratory of Materials Physics, Institute of Solid State Physics, Chinese Academy of Sciences, Hefei, 230031, China}

\author{Wei Gan}
\affiliation{Institute of Physical Science and Information Technology, Anhui University, Hefei 230601, China}
\author{Huan-Wen Wang}
\affiliation{School of Physics, University of Electronic Science and Technology of China, Chengdu 610054, China}  

\author{Hui Han}
\affiliation{Institute of Physical Science and Information Technology, Anhui University, Hefei 230601, China}

\author{Z. G. Sheng}
\affiliation{High Magnetic Field Laboratory, Chinese Academy of Sciences, Hefei 230031, China}

\author{Changjin Zhang}
\affiliation{Institute of Physical Science and Information Technology, Anhui University, Hefei 230601, China}
\affiliation{High Magnetic Field Laboratory, Chinese Academy of Sciences, Hefei 230031, China}

\author{Hua Jiang} \email{jianghuaphy@suda.edu.cn}
\affiliation{Institute for Advanced Study, Soochow University, Suzhou 215006, China}
\affiliation{Institute for Nanoelectronic Devices and Quantum Computing, Fudan University, Shanghai 200433, China}

\author{Hui Li} \email{huili@ahu.edu.cn}
\affiliation{Institute of Physical Science and Information Technology, Anhui University, Hefei 230601, China}

\begin{abstract}
The relationship between {magnetic order} and the second harmonic generation (SHG) effect is a fundamental area of study in condensed matter physics with significant practical implications. {In order to gain a clearer understanding of this intricate relation, this study presents a comprehensive classification scheme for the SHG effect in magnetically ordered materials. This framework offers a straightforward approach to connect magnetic order and SHG effect. The characteristics of the SHG tensors in all magnetic point groups are studied using the isomorphic group method, followed by a comprehensive SHG effect classification scheme that includes seven types based on the symmetries of the magnetic phases and their corresponding parent phases. In addition, a tensor dictionary containing the SHG and linear magneto-optic (LMO) effect is established. Furthermore, an extensive SHG database of magnetically ordered materials is also built up. This classification strategy exposes an anomalous SHG effect with even characteristic under time-reversal symmetry, which is solely contributed by magnetic structure. Moreover, the proposed classification scheme facilitates the determination of magnetic structures through SHG effect.}
\end{abstract}

\maketitle

\noindent \textbf{INTRODUCTION}

Symmetry plays a crucial role in determining the physical properties of matter.
In {magnetically ordered materials (such as ferromagnetic (FM) materials, antiferromagnetic (AFM) materials, ferrimagnetic materials, etc.),} the spin directions of electrons are arranged in orderly fashions, leading to spontaneous time-reversal symmetry breaking and rich physical phenomena.
{To reveal the symmetries of magnetically ordered materials, optical technical is usually considered as an effective approach} \cite{RN2464, RN2179, RN2722}.
The linear magneto-optic (LMO) effect ({refers specifically to} the Faraday effect \cite{RN3615} in transmission and Kerr effect \cite{RN3613} in reflection here) and the second harmonic generation (SHG) effect are two basic and complemental optical tools {to study magnetic structures} in experiments.
SHG effect {is particularly powerful} in characterizing AFM materials where the LMO method fails, particularly in recent two-dimensional (2D) AFM materials, for example, bilayer CrI$_3$ \cite{RN2808}, MnPS$_3$ \cite{RN2518}, MnPSe$_3$ \cite{RN3052}, CrSBr \cite{RN3196} and NiI$_2$ \cite{RN3350}. {Therefore}, SHG  {is widely adopted to} detect magnetic phase transitions \cite{RN3337, RN2843, RN2501, RN2838, RN2521}, magnetic symmetries, magnetic orders \cite{RN3322, RN3336, RN3351} and domain structures \cite{RN2838, RN2501, RN3544}, due to {its} spectral and spatial resolution. Therefore, bridging the connection between the SHG effect and magnetic order {is a frontier area in this field} \cite{RN2843, RN2501}.

Comprehensively classifying the SHG effect in {magnetically ordered materials} offers {a promising approach to better understand the intriguing relationship between the SHG effect and the magnetic structures}.
However, {achieving this goal is still elusive}. It is well-known that the appearance of the SHG effect may indicate the breaking of inversion symmetry under the electric dipole approximation. In {magnetically ordered materials}, the inversion-symmetry-breaking {can} be induced by either asymmetric crystal structures ({inversion symmetry breaking resulting from the parent crystallographic structures}) or magnetic structures \cite{RN2813, RN3393, RN3394, RN2124, RN2842, RN3003, RN3300}.
{Birss, in the 1960s \cite{RN3376, RN3378}, divided SHG tensor elements into two types based on their parities under time-reversal symmetry: $i$-type (even with time-reversal symmetry $T$) and $c$-type (odd with time-reversal symmetry $T$).}
Until now, this two-category classification is widely used \cite{RN3337, RN2843, RN2808, RN2518, RN3052, RN3196}. However, the classification {faces} daunting challenges in
{distinguishing whether the inversion symmetry breaking}.
{Furthermore, the SHG classification is not mutually exclusive and spatial symmetry is not considered, leading to the complexity in characterizing the SHG effect.}
Moreover, the unsystematic study of all magnetic groups {may} lead to some misconceptions. For example, {the SHG effects in magnetically ordered materials are assumed to be odd with time reversal symmetry} \cite{RN2808, RN2518, RN3052, RN3196, RN3350}, since the {magnetic structure} is reversed under the time-reversal operator.
However, {it should be noted that the even SHG tensors can also be induced by magnetic order individually} because no symmetry prohibits them, and this anomalous SHG effect has not been realized in previous literature. Therefore, a comprehensive SHG classification that reveals the characteristics of SHG effect and {accurately identifies the origin of the inversion symmetry breaking, whether it arises from crystallographic or magnetic structures} is highly desirable. 

In this study, we introduce an SHG classification method in magnetically ordered materials containing these two {essential} aspects: the SHG characteristics and the origin {of the inversion symmetry breaking}. Firstly, we use the \emph{isomorphic group method} to transform the SHG tensor under the magnetic point groups into non-magnetic point groups, {thereby revealing the rules governing the SHG tensor}. Based on this, we classify the SHG effect into seven types based on the symmetries of the magnetic phases and parent phases.
Additionally, we build up a tensor dictionary containing the SHG and LMO effects and establish a comprehensive classification for the {magnetically ordered materials} in the MAGNDATA database \cite{RN3334}.
Furthermore, our classification strategy predicts an anomalous SHG effect, which {exhibits even characteristic with time reversal symmetry} and contributed by magnetic structure solely. The first-principles calculations on some representative {magnetically ordered materials} {confirm} the effectiveness of the proposed classification.

\noindent \textbf{RESULTS}

\noindent \textbf{Characteristics of SHG tensors of all magnetic points groups with isomorphic group method}

Typically, SHG susceptibility tensors can be divided into $T$-even ($i$-type, ${\chi}_{ijk}^{\text{even}}$) and $T$-odd ($c$-type, ${\chi}_{ijk}^{\text{odd}}$) parts \cite{RN3376, RN3378},
\begin{equation}
\chi_{ijk}^{(2)} ={{\chi }_{ijk}^{\text{even}}}+{{\chi }_{ijk}^{\text{odd}}}.
\end{equation}
The subscript $i$ ($j$ and $k$) denotes the direction of second-order polarization (fundamental incident) of light. The notation of the SHG tensor is presented in Supplementary Notes 1.1 and 3.2.

According to Neumann's principle \cite{RN3376, RN3384, RN2905, RN2165}, the SHG tensors are constrained by the magnetic point groups (MPGs, {denoted as} $M_0$) rather than the magnetic space groups (MSGs, denoted as $M$, {also known as} Shubnikov groups) \cite{RN3388, RN1516, RN3398}.
{
The point group operations in MPGs contain two categories: the unitary point operators $R$, which do not involve the time operation, such as rotation operation $n$ and rotation-inversion operation $\bar{n}$ ($n$=1, 2, 3, 4, 6), and the anti-unitary point operation $R^{\prime}$. The relationship between $R$ and $R^{\prime}$ is described as $R^{\prime}= TR$, where $T$ is the time reversal symmetry.}
The MPGs can be divided into three classes {based on the presence of the $R$ and $R'$}: (a) original MPGs, {which are constituted by ordinary point groups $G_0$ and lack} any anti-unitary operators $R'$, i.e. $M_0=G_0$, (b) grey MPGs, {in the from of $M_0=G_0 + TG_0=G_0+1'G_0$ ($1'=1T$ also means the $T$ symmetry)}, (c) black-white (BW) MPGs  {$M_0=S_0 + T(G_0-S_0)=S_0+ R'S_0$, where $S_0$ is a halving subgroup of $G_0$}.
A brief introduction to MPGs and MSGs is {provided} in Supplementary Note 1.2.


The transformations of the even and odd SHG tensors under a unitary point operation $R$ {can be expressed as follows}:
\begin{equation}
R:\chi_{ijk}^{\text{even}/\text{odd}}=\sum\limits_{lmn}{{{R}_{il}}}{{R}_{jm}}{{R}_{kn}}\chi_{lmn}^{\text{even}/\text{odd}},
\label{Eq-sym_chi_R}
\end{equation}
where $R_{ij}$ {represents} an element of the 3$\times$3 matrix {for} the unitary point operation $R$. According to Eq. (\ref{Eq-sym_chi_R}), the transformations of the {even and odd SHG tensors, i.e., $\chi_{ijk}^{\text{even}}$ and $\chi_{ijk}^{\text{odd}}$, are identical} under $R$, since $R$ does not contain the time-reversal operation $T$. However, {the signs of} $\chi_{ijk}^{\text{even}}$ and $\chi_{ijk}^{\text{odd}}$ {are opposite under the transformations of} the anti-unitary point operation $R'$:
\begin{equation}
{R}':\chi_{ijk}^{\text{even}/\text{odd}}=\pm \sum\limits_{lmn}{{{R}_{il}}}{{R}_{jm}}{{R}_{kn}}\chi_{lmn}^{\text{even}/\text{odd}}.
\label{Eq-sym_chi_R'}
\end{equation}
The plus and minus sign in Eq. (\ref{Eq-sym_chi_R'}) correspond to the even part ($\chi_{ijk}^{\text{even}}$) and the odd part ($\chi_{ijk}^{\text{odd}}$) of SHG tensors, respectively \cite{RN2905}.

Generally, the {characteristics} of SHG tensors under all MPGs can be directly obtained by solving the above linear algebra equations with the {aid of} invariant theory (see Refs. \cite{RN3376, RN3384, RN2905} and \href{https://www.cryst.ehu.es/cgi-bin/cryst/programs/mtensor.pl}{``Bilbao Crystallographic Server/MTENSOR"} website). {However}, the SHG tensor characteristics in {magnetically ordered materials} are more complicated than that of non-magnetically ordered  materials. Currently, the relationships between MPGs and SHG tensors as well as the rules governing the SHG effect under all MPGs have not been {clearly established}. This lack of understanding hinders the study of the SHG effect.

Here, we {investigate} the SHG tensors using the \emph{isomorphic group method}. {In group theory}, $-1$ is {equivalent} to the inversion operator $I$, because $-1=-{{({{I}_{0}})}_{3\times 3}}$ (${{I}_{0}}$ is the identity matrix) and {$-{{I}_{0}}$ is the matrix of the inversion operation}. Therefore, the transformation of $\chi_{ijk}^{\text{odd}}$ in Eq. (\ref{Eq-sym_chi_R'}) under the anti-unitary point operation $R'$ can be re-written as {follows}:
\begin{equation}
\begin{aligned}
& {R}':\chi_{ijk}^{\text{odd}}=-\sum\limits_{lmn}{{{R}_{il}}}{{R}_{jm}}{{R}_{kn}}\chi_{lmn}^{\text{odd}} \\
& =\sum\limits_{lmn}{{{\left( -R \right)}_{il}}}{{\left( -R \right)}_{jm}}{{\left( -R \right)}_{kn}}\chi_{lmn}^{\text{odd}} \\
& =\sum\limits_{lmn}{{{\left( IR \right)}_{il}}}{{\left( IR \right)}_{jm}}{{\left( IR \right)}_{kn}}\chi_{lmn}^{\text{odd}}. \\
\label{Eq-sym_chi_R'-T}
\end{aligned}
\end{equation}
The above equation {implies} that the constraint {imposed by} the anti-unitary point operation $R'$ on $\chi_{ijk}^{\text{odd}}$ is equivalent to an unitary {point operation $R$ times an inversion operation $I$, i.e. $IR$}.

{The set of} all the unitary $R$ and anti-unitary $R'$ operations in a specific magnetically ordered material constitute a closed MPG.
The transformations of $\chi_{ijk}^{\text{odd}}$ under all the unitary point operations $R$ ($ R \in S_0$) and anti-unitary point operations $R'$ ($R' \in R'S_0$) of a BW MPG $M_0$ are equivalent to $S_0$ and $IRS_0$, respectively. {Thus}, the symmetry restrictions {imposed by} a BW MPG $M_0$ on $\chi_{ijk}^{\text{odd}}$ is equivalent to a nonmagnetic point group (PG) $G_1$ ($= S_0+IRS_0$).
Similarly, the symmetry constraint of $\chi_{ijk}^{\text{even}}$ {imposed by} a BW MPG $M_0$ is equivalent to a nonmagnetic PG $G_0$ ($= S_0+ RS_0$), because the transformations of $\chi_{ijk}^{\text{even}}$ under $R$ and $R'$ are {identical}, {as indicated in} Eq. (\ref{Eq-sym_chi_R}) and Eq. (\ref{Eq-sym_chi_R'}). Because $M_0$, $G_0$ and $G_1$ are \emph{isomorphic groups}, {this implies} the isomorphic group method can be {employed} to transform the SHG tensor under MPGs into nonmagnetic PGs.

\begin{figure}[!htbp]
\includegraphics[width=0.48\textwidth]{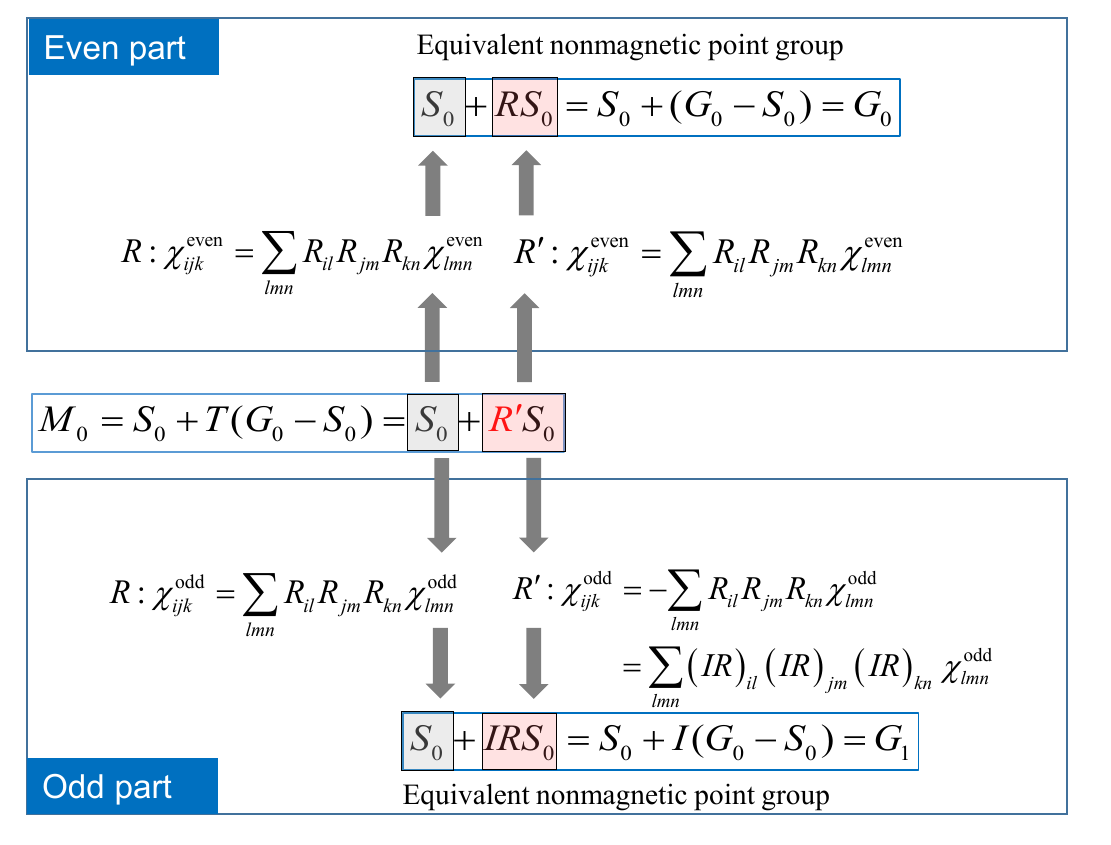}
\caption{\textbf{Diagram of the isomorphic group method to analyze the {characteristic} of SHG tensor.} The transformations of $\chi_{ijk}^{\text{even}}$ and $\chi_{ijk}^{\text{odd}}$ under {a BW} MPG $M_0$ ($=S_0+R'S_0$) are equivalent to {those under} nonmagnetic PGs $G_0$ ($=S_0+RS_0$, top panel) and $G_1$ ($=S_0+IRS_0$, bottom panel), respectively. $S_0$ is a halving subgroup of MPG $M_0$. $R$ is a unitary point operation, $R'$ is {an} anti-unitary point operation, and $I$ is the inversion operation. {Further information is available in Supplementary Note 1.}}
\label{Fig-Isomorphic_group_method_scheme}
\end{figure}

As {depicted} in Fig. \ref{Fig-Isomorphic_group_method_scheme}, we {utilize} two nonmagnetic isomorphic PGs, $G_0$ and $G_1$, to {determine} the characteristics of the even ($\chi_{ijk}^{\text{even}}$) and odd ($\chi_{ijk}^{\text{odd}}$) SHG tensors under {a BW} MPG $M_0$, respectively.
{The SHG tensors of grey and original MPGs can also obtained using the above isomorphic group method, which are given in Supplementary Note 1.3}.
No additional numerical calculations {are needed}, since the characteristics of the SHG tensors of {all} the nonmagnetic PGs can be {found} in general nonlinear optics textbooks \cite{Boyd2010, Shi2012}.
{Importantly}, the isomorphic group method {elucidates the rules governing} the SHG tensors \cite{RN3376, RN3384, RN2905}.

{The characteristics of LMO effect in all MPGs are also studied in Supplementary Note 2. In our work, LMO effect specifically refers to the Faraday effect \cite{RN3615} and Kerr effect \cite{RN3613}, both of which belong to the linear optics.
These two kinds of LMO effects arise from the magneto-circular dichroism in transmission (Faraday effect) or reflectivity (Kerr effect) of polarized light in magnetically ordered materials. It was once believed that LMO effect could only exist in magnetically ordered materials with nonzero net magnetization.
However, with the the permitted symmetries, LMO effect has been theoretically predicted or experimentally observed in AFM materials with net magnetization \cite{RN3536}. Specific examples include collinear AFM materials (such as RuO$_2$ \cite{RN3505, RN3506, RN3466} and MnTe \cite{RN3594}), coplanar but non-collinear AFM materials (such as Mn$_3$X (X = Rh, Ir, Pt) \cite{RN3181}, Mn$_3$Y (Y = Ge, Ga, Sn) \cite{RN2479, RN3609, RN3608} and Mn$_3$ZN (Z = Ga, Zn, Ag, Ni) \cite{RN3616}), as well as in non-coplanar AFM materials (such as $\gamma$-Fe$_{0.5}$Mn$_{0.5}$ \cite{RN2475} and K$_{0.5}$RhO$_{2}$ \cite{RN2475}).}
In all, LMO effect can exist in magnetically ordered materials with permitted MPGs.
{The higher order magneto-optic effects, such as spontaneous nonreciprocal optical effect \cite{RN3602, RN3610, RN2849, RN2850} and the magneto-birefringence effect \cite{RN2880} (Voigt effect \cite{RN3611} and Cotton-Mouton effect \cite{RN3612}), are not investigated in our work.}

As {stated in} the introduction, the LMO and SHG effects are two complementary optical methods to study the {magnetically ordered materials}. The {characteristics} of SHG and LMO effects in all MPGs are listed in Supplementary Note 3, {which provides} a dictionary for characterizing the magnetic structures with these two optical techniques in experiments. Moreover, this isomorphic group method is {also suitable} for other tensors in magnetic systems, such as circular photo-galvanic effect (CPGE) whose susceptibilities {comprise} 3$\times$3 pseudo-tensors (see Supplementary Note 4 for detail). This indicates that this method is a powerful {tool} for studying the tensors in magnetic systems.

\begin{table*}[!htpb]
\centering
\caption{\textbf{Characteristics of SHG tensors in all MPGs {without inversion symmetry}.}}
\label{Tab2-MPG_SHG}
\begin{threeparttable}
\begin{tabular} {m{3cm}|c|c|m{6cm}} 
\hline
\hline
\textbf{\quad\quad SHG cases} & \textbf{Even SHG tensor} & \textbf{Odd SHG tensor} & \textbf{ \quad\quad\quad\quad\quad\quad MPGs} \\
\hline
Case 1. Original MPG without $P$ (20) &
\begin{tabular}{c}
$\checkmark$ \\ ${{\chi}_{ijk}^{\text{even}}}\left({{G}_{0}}\right)$ \\ Same as odd
\end{tabular}
&	
\begin{tabular}{c}
$\checkmark$ \\ ${{\chi }_{ijk}^{\text{odd}}}\left({{G}_{0}}\right)$ \\ Same as even
\end{tabular}
&
\href{https://www.cryst.ehu.es/cgi-bin/cryst/programs/nph-mpoint?magnum=1.1.1\&way=main} {$1$}(M),
\href{https://www.cryst.ehu.es/cgi-bin/cryst/programs/nph-mpoint?magnum=3.1.6\&way=main}	{$2$}(M),
\href{https://www.cryst.ehu.es/cgi-bin/cryst/programs/nph-mpoint?magnum=4.1.9\&way=main}	{$m$}(M),
\href{https://www.cryst.ehu.es/cgi-bin/cryst/programs/nph-mpoint?magnum=6.1.17\&way=main}	{$222$},
\href{https://www.cryst.ehu.es/cgi-bin/cryst/programs/nph-mpoint?magnum=7.1.20\&way=main}	{$mm2$},
\href{https://www.cryst.ehu.es/cgi-bin/cryst/programs/nph-mpoint?magnum=9.1.29\&way=main}	{$4$}(M),
\href{https://www.cryst.ehu.es/cgi-bin/cryst/programs/nph-mpoint?magnum=10.1.32\&way=main}	{$\bar{4}$}(M),
\href{https://www.cryst.ehu.es/cgi-bin/cryst/programs/nph-mpoint?magnum=12.1.40\&way=main}	{$422$},
\href{https://www.cryst.ehu.es/cgi-bin/cryst/programs/nph-mpoint?magnum=13.1.44\&way=main}	{$4mm$},
\href{https://www.cryst.ehu.es/cgi-bin/cryst/programs/nph-mpoint?magnum=14.1.48\&way=main}	{$\bar{4}2m$},
\href{https://www.cryst.ehu.es/cgi-bin/cryst/programs/nph-mpoint?magnum=16.1.60\&way=main}	{$3$}(M),
\href{https://www.cryst.ehu.es/cgi-bin/cryst/programs/nph-mpoint?magnum=18.1.65\&way=main}	{$32$},
\href{https://www.cryst.ehu.es/cgi-bin/cryst/programs/nph-mpoint?magnum=19.1.68\&way=main}	{$3m$},
\href{https://www.cryst.ehu.es/cgi-bin/cryst/programs/nph-mpoint?magnum=21.1.76\&way=main}	{$6$}(M),
\href{https://www.cryst.ehu.es/cgi-bin/cryst/programs/nph-mpoint?magnum=22.1.79\&way=main}	{$\bar{6}$}(M),
\href{https://www.cryst.ehu.es/cgi-bin/cryst/programs/nph-mpoint?magnum=24.1.87\&way=main}	{$622$},
\href{https://www.cryst.ehu.es/cgi-bin/cryst/programs/nph-mpoint?magnum=25.1.91\&way=main}	{$6mm$},
\href{https://www.cryst.ehu.es/cgi-bin/cryst/programs/nph-mpoint?magnum=26.1.95\&way=main}	{$\bar{6}m2$},
\href{https://www.cryst.ehu.es/cgi-bin/cryst/programs/nph-mpoint?magnum=28.1.107\&way=main}	{$23$},
\href{https://www.cryst.ehu.es/cgi-bin/cryst/programs/nph-mpoint?magnum=31.1.115\&way=main}	{$\bar{4}3m$}
\\ \hline

Case 2. Grey MPG without $P$ (20)	&
\begin{tabular}{c}
$\checkmark$ \\ ${{\chi }_{ijk}^{\text{even}}}\left( {{G}_{0}}\right)$	
\end{tabular}
& $\times$ &
\href{https://www.cryst.ehu.es/cgi-bin/cryst/programs/nph-mpoint?magnum=1.2.2\&way=main}	{$11'$},
\href{https://www.cryst.ehu.es/cgi-bin/cryst/programs/nph-mpoint?magnum=3.2.7\&way=main}	{$21'$},
\href{https://www.cryst.ehu.es/cgi-bin/cryst/programs/nph-mpoint?magnum=4.2.10\&way=main}	{$m1'$},
\href{https://www.cryst.ehu.es/cgi-bin/cryst/programs/nph-mpoint?magnum=6.2.18\&way=main}	{$2221'$},
\href{https://www.cryst.ehu.es/cgi-bin/cryst/programs/nph-mpoint?magnum=7.2.21\&way=main}	{$mm21'$},
\href{https://www.cryst.ehu.es/cgi-bin/cryst/programs/nph-mpoint?magnum=9.2.30\&way=main}	{$41'$},
\href{https://www.cryst.ehu.es/cgi-bin/cryst/programs/nph-mpoint?magnum=10.2.33\&way=main}	{$\bar{4}1'$},
\href{https://www.cryst.ehu.es/cgi-bin/cryst/programs/nph-mpoint?magnum=12.2.41\&way=main}	{$4221'$},
\href{https://www.cryst.ehu.es/cgi-bin/cryst/programs/nph-mpoint?magnum=13.2.45\&way=main}	{$4mm1'$},
\href{https://www.cryst.ehu.es/cgi-bin/cryst/programs/nph-mpoint?magnum=14.2.49\&way=main}	{$\bar{4}2m1'$},
\href{https://www.cryst.ehu.es/cgi-bin/cryst/programs/nph-mpoint?magnum=16.2.61\&way=main}	{$31'$},
\href{https://www.cryst.ehu.es/cgi-bin/cryst/programs/nph-mpoint?magnum=18.2.66\&way=main}	{$321'$},
\href{https://www.cryst.ehu.es/cgi-bin/cryst/programs/nph-mpoint?magnum=19.2.69\&way=main}	{$3m1'$},
\href{https://www.cryst.ehu.es/cgi-bin/cryst/programs/nph-mpoint?magnum=21.2.77\&way=main}	{$61'$},
\href{https://www.cryst.ehu.es/cgi-bin/cryst/programs/nph-mpoint?magnum=22.2.80\&way=main}	{$\bar{6}1'$},
\href{https://www.cryst.ehu.es/cgi-bin/cryst/programs/nph-mpoint?magnum=24.2.88\&way=main}	{$6221'$},
\href{https://www.cryst.ehu.es/cgi-bin/cryst/programs/nph-mpoint?magnum=25.2.92\&way=main}	{$6mm1'$},
\href{https://www.cryst.ehu.es/cgi-bin/cryst/programs/nph-mpoint?magnum=26.2.96\&way=main}	{$\bar{6}m21'$},
\href{https://www.cryst.ehu.es/cgi-bin/cryst/programs/nph-mpoint?magnum=28.2.108\&way=main}	{$231'$},
\href{https://www.cryst.ehu.es/cgi-bin/cryst/programs/nph-mpoint?magnum=31.2.116\&way=main}	{$\bar{4}3m1'$}
\\ \hline
Case 3. BW MPG without $P$ but with $PT$ (20) &	$\times$	&
\begin{tabular}{c}
$\checkmark$ \\ ${{\chi }_{ijk}^{\text{odd}}}\left( {{S}_{0}} \right)$
\end{tabular}
&
\href{https://www.cryst.ehu.es/cgi-bin/cryst/programs/nph-mpoint?magnum=2.3.5\&way=main}	{$\bar{1}'$},
\href{https://www.cryst.ehu.es/cgi-bin/cryst/programs/nph-mpoint?magnum=5.3.14&way=main}	{$2'/m$},
\href{https://www.cryst.ehu.es/cgi-bin/cryst/programs/nph-mpoint?magnum=5.4.15&way=main} 	{$2/m'$},
\href{https://www.cryst.ehu.es/cgi-bin/cryst/programs/nph-mpoint?magnum=8.5.28\&way=main}	{$m'm'm'$},
\href{https://www.cryst.ehu.es/cgi-bin/cryst/programs/nph-mpoint?magnum=8.3.26\&way=main}	{$mmm'$},
\href{https://www.cryst.ehu.es/cgi-bin/cryst/programs/nph-mpoint?magnum=11.4.38\&way=main}	{$4/m'$},
\href{https://www.cryst.ehu.es/cgi-bin/cryst/programs/nph-mpoint?magnum=11.5.39\&way=main}	{$4'/m'$},
\href{https://www.cryst.ehu.es/cgi-bin/cryst/programs/nph-mpoint?magnum=15.3.55\&way=main}	{$4/m'mm$},
\href{https://www.cryst.ehu.es/cgi-bin/cryst/programs/nph-mpoint?magnum=15.5.57\&way=main}	{$4'/m'm'm$},
\href{https://www.cryst.ehu.es/cgi-bin/cryst/programs/nph-mpoint?magnum=15.7.59\&way=main}	{$4/m'm'm'$},
\href{https://www.cryst.ehu.es/cgi-bin/cryst/programs/nph-mpoint?magnum=17.3.64\&way=main}	{$\bar{3}'$},
\href{https://www.cryst.ehu.es/cgi-bin/cryst/programs/nph-mpoint?magnum=20.3.73\&way=main}	{$\bar{3}'m$},
\href{https://www.cryst.ehu.es/cgi-bin/cryst/programs/nph-mpoint?magnum=20.4.74\&way=main}	{$\bar{3}'m'$},
\href{https://www.cryst.ehu.es/cgi-bin/cryst/programs/nph-mpoint?magnum=23.3.84\&way=main}	{$6'/m$},
\href{https://www.cryst.ehu.es/cgi-bin/cryst/programs/nph-mpoint?magnum=23.4.85\&way=main}	{$6/m'$},
\href{https://www.cryst.ehu.es/cgi-bin/cryst/programs/nph-mpoint?magnum=27.3.102\&way=main}	{$6/m'mm$},
\href{https://www.cryst.ehu.es/cgi-bin/cryst/programs/nph-mpoint?magnum=27.4.103\&way=main}	{$6'/mmm'$},
\href{https://www.cryst.ehu.es/cgi-bin/cryst/programs/nph-mpoint?magnum=27.7.106\&way=main}	{$6/m'm'm'$},
\href{https://www.cryst.ehu.es/cgi-bin/cryst/programs/nph-mpoint?magnum=29.3.111\&way=main}	{$m'\bar{3}'$},
\href{https://www.cryst.ehu.es/cgi-bin/cryst/programs/nph-mpoint?magnum=32.3.120\&way=main}	{$m'\bar{3}'m$}
\\ \hline

Case 4. BW MPG without $P$ nor $PT$ (27)	&
\begin{tabular}{c}$\checkmark$ \\ ${{\chi }_{ijk}^{\text{even}}}\left( {{G}_{0}}\right)$ \\Different with odd
\end{tabular}&
\begin{tabular}{c}$\checkmark$ \\ ${{\chi }_{ijk}^{\text{odd}}}\left( {{G}_{1}}\right)$ \\ Different with even
\end{tabular}
&
\href{https://www.cryst.ehu.es/cgi-bin/cryst/programs/nph-mpoint?magnum=3.3.8\&way=main}	{$2'$}(M),
\href{https://www.cryst.ehu.es/cgi-bin/cryst/programs/nph-mpoint?magnum=4.3.11&way=main} 	{$m'$} (M),
\href{https://www.cryst.ehu.es/cgi-bin/cryst/programs/nph-mpoint?magnum=6.3.19\&way=main}	{$2'2'2$}(M),
\href{https://www.cryst.ehu.es/cgi-bin/cryst/programs/nph-mpoint?magnum=7.4.23&way=main} 	{$m'm'2$} (M),
\href{https://www.cryst.ehu.es/cgi-bin/cryst/programs/nph-mpoint?magnum=7.3.22&way=main}	{$m'm2'$} (M),
\href{https://www.cryst.ehu.es/cgi-bin/cryst/programs/nph-mpoint?magnum=9.3.31\&way=main}	{$4'$},
\href{https://www.cryst.ehu.es/cgi-bin/cryst/programs/nph-mpoint?magnum=10.3.34\&way=main}	{$\bar{4}'$},
\href{https://www.cryst.ehu.es/cgi-bin/cryst/programs/nph-mpoint?magnum=12.3.42\&way=main}	{$4'22'$},
\href{https://www.cryst.ehu.es/cgi-bin/cryst/programs/nph-mpoint?magnum=12.4.43\&way=main}	{$42'2'$}(M),
\href{https://www.cryst.ehu.es/cgi-bin/cryst/programs/nph-mpoint?magnum=13.3.46\&way=main}	{$4'm'm$},
\href{https://www.cryst.ehu.es/cgi-bin/cryst/programs/nph-mpoint?magnum=13.4.47\&way=main}	{$4m'm'$}(M),
\href{https://www.cryst.ehu.es/cgi-bin/cryst/programs/nph-mpoint?magnum=14.3.50\&way=main}	{$\bar{4}'2'm$},
\href{https://www.cryst.ehu.es/cgi-bin/cryst/programs/nph-mpoint?magnum=14.4.51\&way=main}	{$\bar{4}'2m'$},
\href{https://www.cryst.ehu.es/cgi-bin/cryst/programs/nph-mpoint?magnum=14.5.52\&way=main}	{$\bar{4}2'm'$}(M),
\href{https://www.cryst.ehu.es/cgi-bin/cryst/programs/nph-mpoint?magnum=18.3.67\&way=main}	{$32'$}(M),
\href{https://www.cryst.ehu.es/cgi-bin/cryst/programs/nph-mpoint?magnum=19.3.70\&way=main}	{$3m'$}(M),
\href{https://www.cryst.ehu.es/cgi-bin/cryst/programs/nph-mpoint?magnum=21.3.78\&way=main}	{$6'$},
\href{https://www.cryst.ehu.es/cgi-bin/cryst/programs/nph-mpoint?magnum=22.3.81\&way=main}	{$\bar{6}'$},
\href{https://www.cryst.ehu.es/cgi-bin/cryst/programs/nph-mpoint?magnum=24.3.89\&way=main}	{$6'22'$},
\href{https://www.cryst.ehu.es/cgi-bin/cryst/programs/nph-mpoint?magnum=24.4.90\&way=main}	{$62'2'$}(M),
\href{https://www.cryst.ehu.es/cgi-bin/cryst/programs/nph-mpoint?magnum=25.3.93\&way=main}	{$6'mm'$},
\href{https://www.cryst.ehu.es/cgi-bin/cryst/programs/nph-mpoint?magnum=25.4.94\&way=main}	{$6m'm'$}(M),
\href{https://www.cryst.ehu.es/cgi-bin/cryst/programs/nph-mpoint?magnum=26.3.97\&way=main}	{$\bar{6}'m'2$},
\href{https://www.cryst.ehu.es/cgi-bin/cryst/programs/nph-mpoint?magnum=26.4.98\&way=main}	{$\bar{6}'m2'$},
\href{https://www.cryst.ehu.es/cgi-bin/cryst/programs/nph-mpoint?magnum=26.5.99&way=main}{$\bar{6}m'2'$}(M),
\href{https://www.cryst.ehu.es/cgi-bin/cryst/programs/nph-mpoint?magnum=30.3.114&way=main}	{$4'32'$},
\href{https://www.cryst.ehu.es/cgi-bin/cryst/programs/nph-mpoint?magnum=31.3.117&way=main}	{$\bar{4}'3m'$}
\\ \hline
\hline
\end{tabular}

\begin{tablenotes}
 \footnotesize
 \item[1] The corresponding SHG tensors are presented in Supplementary Note 3.
 \item[2] The numbers in the first column {represent} the number of MPGs {exhibiting} the SHG effect.
 \item[3] {The letter} ``M" in the bracket of the fourth column indicates the MPG {permitting} the LMO effect.
 \end{tablenotes}
\end{threeparttable}

\end{table*}

\begin{figure*}[!htbp]
\centering
\includegraphics[width=0.9\textwidth]{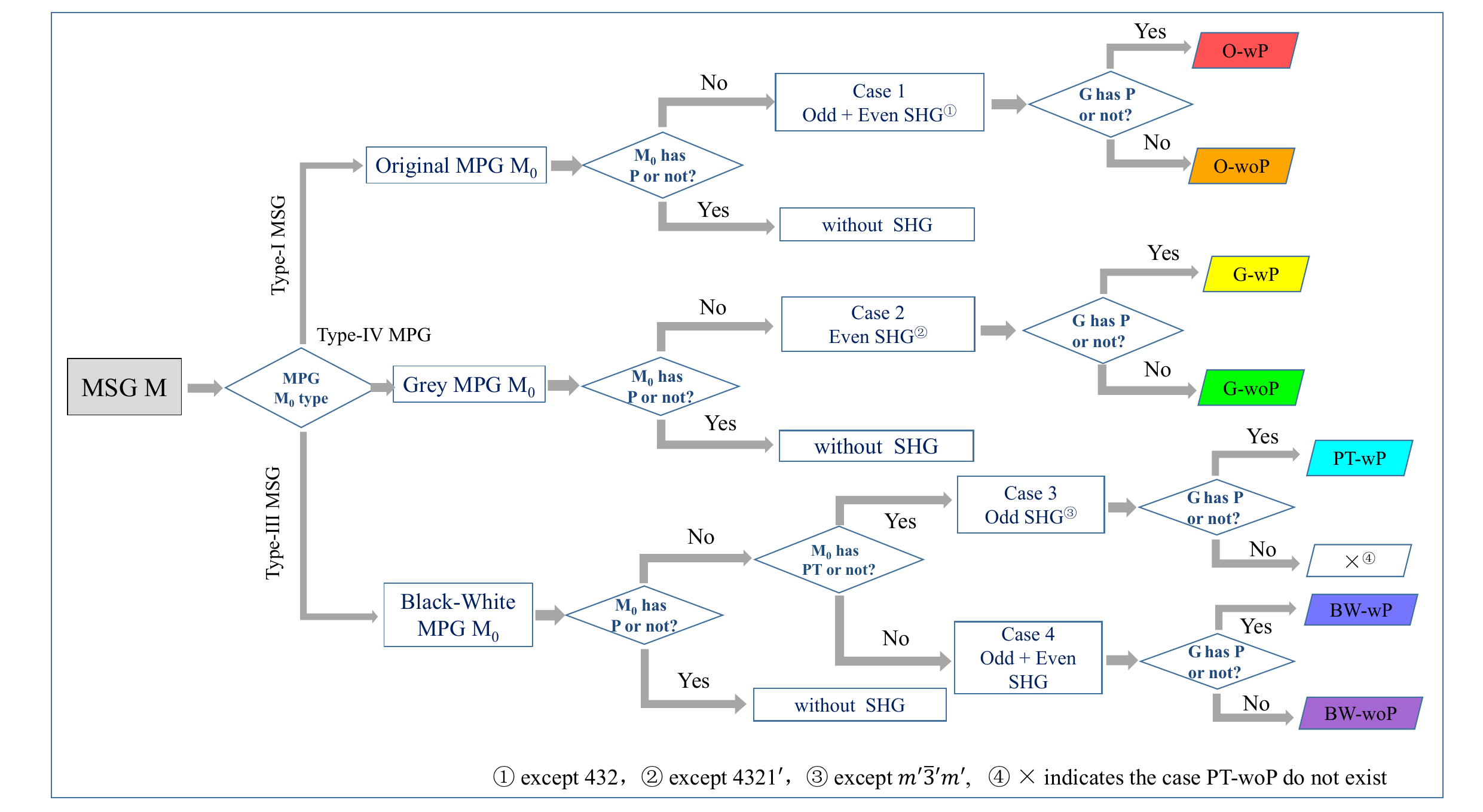}
\caption{\textbf{Flow chart for {SHG classification scheme} in {magnetically ordered materials}.} First, the MPGs are determined from the MSGs of {magnetically ordered materials}. Then, the SHG tensors are further divided into four cases based on the symmetries of MPGs (refer to Table \ref{Tab2-MPG_SHG}). Finally, the types of SHG effect in {magnetically ordered materials} are determined based on whether the parent phases possess inversion symmetry (wP) or not (woP). {In this notation,} $M$ represents the MSG, and $G$ in the last judging rhombus denotes the space group (SG) of the parent phase. The colored parallelograms on the right {indicate} the types of SHG effect.
{The letter ``G" in G-wP and G-woP denotes the grey MPG, and the letter ``O" in O-wP and O-woP represents the original MPG. Similarly, ``BW" in BW-wP and BW-woP represents the BW MPG without $PT$ symmetry, and ``PT" in PT-wP indicates the BW MPG with $PT$ symmetry.}
{Note that} MPGs of 432, 4321$^{\prime}$ and {$m^{\prime} \overline{3}^{\prime} m^{\prime}$} are excluded because their symmetries forbid the emerge of SHG effect, even though inversion symmetry is broken.}

\label{Fig3-Flow_chart}
\end{figure*}

According to our symmetry analysis, the SHG tensors in all MPGs can be divided into four cases based on {their characteristics}: (1) original MPGs without inversion symmetry, (2) grey MPGs without inversion symmetry, (3) BW MPGs without inversion symmetry but with $PT$ symmetry (combination of time-reversal $T$ and spatial inversion $P$ symmetry), and (4) BW MPGs without inversion symmetry nor $PT$ symmetry. The MPGs and the corresponding SHG characteristics are listed in Table \ref{Tab2-MPG_SHG}. Only even SHG tensors exist in grey MPGs (Case 2). Only the odd SHG tensors exist in BW MPGs with $PT$ symmetry (Case 3).
{In original MPGs without inversion symmetry} (Case 1) and BW MPGs without $P$ or $PT$ symmetry (Case 4), both the odd and even SHG tensors exist. $\chi_{ijk}^{\text{even}}$ and $\chi_{ijk}^{\text{odd}}$ are the same in Case 1. However,  $\chi_{ijk}^{\text{even}}$ and $\chi_{ijk}^{\text{odd}}$ {differ} in Case 4 as they are constrained by different {isomorphic} PGs ($G_0$ and $G_1$). Table \ref{Tab2-MPG_SHG} provides a clear and concise summary of the SHG behavior for different MPGs, which reveals the rules governing the SHG effect under all MPGs.


\noindent \textbf{Classification of the SHG effect in {magnetically ordered materials}}

Next, we classify the SHG effect in {magnetically ordered materials} based on the symmetries of magnetic and parent structures to reveal the characteristics of SHG effect further. {In view of the fact} that breaking the inversion symmetry could {arise from either} crystallographic or magnetic structure, we should also take the symmetry of the parent phase into account {in addition to} the magnetic phase, similar to the classification in species of ferromagnetic, ferroelectric and ferroelastic crystals \cite{RN3547}.
{In magnetically ordered materials, the parent phases usually exhibit higher symmetry than the magnetic phases. As a consequence, the inversion symmetry broken in the magnetic phases could still be retained in the parent phases.}
{Therefore}, each case in Table \ref{Tab2-MPG_SHG} could be further {divided} into two subcases, depending on whether parent phases possess inversion symmetry (wP, i.e. with inversion symmetry) or not (woP, i.e. without inversion symmetry). An exception {arises} in Case 3, as the parent phases must {possess} the inversion symmetry $P$ if the MPGs {exhibit} $PT$ symmetry. Figure \ref{Fig3-Flow_chart} {depicts} the flow chart of the {SHG classification scheme}. Table \ref{Tab3-ClassificationSHG}  {presents} the seven types of SHG effect, labeled as O-wP, O-woP, G-wP, G-woP, PT-wP, BW-wP, and BW-woP, {along with} their corresponding characteristics. {Because of encompassing all possible SHG cases in magnetically ordered materials}, this classification is exhaustive and mutually exclusive.

\begin{table*}[!htbp]
\centering
\caption{\textbf{Classification of the SHG effect based on the symmetries of magnetic structures (i.e. MPGs and MSGs) and parent structures (i.e. parent SGs).}}
\label{Tab3-ClassificationSHG}
\begin{threeparttable}
\begin{tabular}{c|c|c} 
\hline \hline

\diagbox{MPGs / MSGs}{Parent SGs}
& With inversion symmetry (wP) 	& Without inversion symmetry (woP) \\ \hline
\begin{tabular}{c} Case 1.	Original MPGs (O)  \\ { without inversion symmetry} \\ (Type-I MSGs) \end{tabular}
&  \begin{tabular}{c} O-wP\\ Even/odd: from magnetic structure  \end{tabular} &	
\begin{tabular}{c} O-woP\\ Even: from magnetic structure + crystal asymmetry \\ Odd: from magnetic structure \end{tabular}
\\ \hline

\begin{tabular}{c} Case 2. Grey MPGs (G) \\ { without inversion symmetry}  \\ (Type-IV MSGs) \end{tabular} &
\begin{tabular}{c} G-wP\\ Even: from magnetic structure \end{tabular} &
\begin{tabular}{c} G-woP\\ Even: from magnetic structure + crystal asymmetry \end{tabular}
\\ \hline

\begin{tabular}{c} Case 3. BW MPGs with \\ $PT$ symmetry (PT) \\ (Type-III MSGs) \end{tabular}&
\begin{tabular}{c} PT-wP \\ Odd: from magnetic structure \end{tabular} &
$\times$ Not possible
\\ \hline

\begin{tabular}{c} Case 4. BW MPGs without $PT$ \\ nor $P$ symmetry (BW) \\ (Type-III MSGs) \end{tabular} &	
\begin{tabular}{c} BW-wP\\ Even/odd: from magnetic structure \end{tabular} 	&
\begin{tabular}{c} BW-woP\\ Even: from magnetic structure + crystal asymmetry \\ Odd: from magnetic structure \end{tabular}
\\
\hline \hline
\end{tabular}
\begin{tablenotes}
 \footnotesize
 \item[1] A brief introduction of MPGs and MSGs is provided in Supplementary Note 1.2.
 \item[2] The type-II MSGs describe the non{magnetically ordered materials with time-reversal symmetry}, {which are beyond the scope of our study, and therefore are not included.
 \item[3] {``Crystal asymmetry" refers to inversion symmetry breaking resulting from the parent crystal structure.}
     }
 \end{tablenotes}
 \end{threeparttable}
\end{table*}

Figure \ref{Fig3-Flow_chart} and Table \ref{Tab3-ClassificationSHG} present a {straightforward but practical classification approach}.  {Additionally}, the {outcomes} presented in Table \ref{Tab3-ClassificationSHG} {offer} valuable insights into the {distinct roles} of magnetic structures and crystallographic structures in SHG effect. In other words, this classification strategy reveals the {symmetries (containing crystallographic and magnetic symmetries) and physical mechanisms underlying} the SHG effect in {magnetically ordered materials}. For example,  {magnetically ordered materials with inversion symmetry in their parent phases}
{exhibit} four distinctive types of SHG effects, namely, O-wP, G-wP, PT-wP and BW-wP. All these types of SHG effects {arise} from magnetic structure rather than crystal asymmetry. {In particular}, {the G-wP type SHG effect is solely contributed by magnetic structures, and only even SHG tensors are present, which has not yet been explored before}.
In contrast, only three possible SHG types are {anticipated} in {magnetically ordered materials without inversion symmetry in their parent phases}, {namely} O-woP, G-woP and BW-woP. The even SHG tensors of these three SHG types {arise} from crystal and magnetic structures, while the odd {SHG tensors solely} originate from magnetic structure. This comprehensive classification is superior to the two-category classification ($c$-type and $i$-type) \cite{RN3376, RN3378}, since {it provides a clearer understanding of the relationship between symmetry and SHG effect}.

\begin{figure}[!htbp]
\centering
\includegraphics[width=0.4\textwidth]{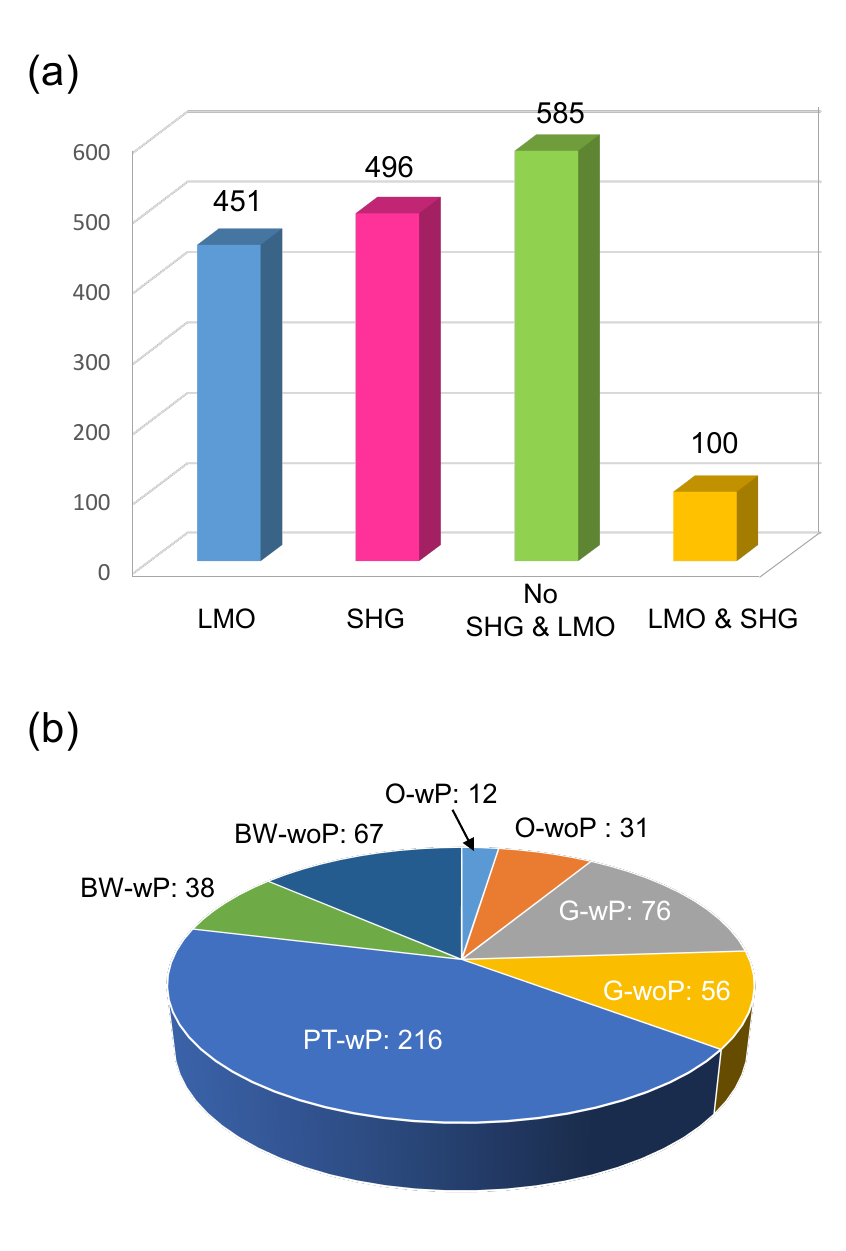}
\caption{\textbf{Statistics on the SHG types in the MAGNDATA database.} (\textbf{a}) Counting of materials with SHG and LMO effects in the 1432 magnetic structures (removing duplicate data). (\textbf{b}) Classification of the 496 magnetic structures with SHG effect. The detailed information of every material is presented in Supplementary Note 6.}
\label{Fig4-MagnDat}
\end{figure}

{Now we apply this classification method to real materials.}
\href{http://webbdcrista1.ehu.es/magndata/}{MAGNDATA} \cite{RN3334} {in the Bilbao Crystallographic Server (BCS)} is a well-known material database that {encompasses over} 1795 magnetic structures.
{It is worth noting that we refer them to magnetic structures}, since one material may possess  several magnetic structures.
{After excluding} the incommensurate magnetic structures (the magnetic lattice constants are not {integer multiples} of those of parent phases)\cite{RN3565}, there are 1655 magnetic structures {remaining} with BCS-ID {(identity number in MAGNDATA of BCS)} {ranging from} 0.1-0.835, 1.0.1-1.0.52, 1.1-1.663, 2.1-2.86 and 3.1-3.19. The corresponding materials are listed in Supplementary Table XI. After removing duplicate data, there are 1432 magnetic structures left. Using the process presented in Fig. \ref{Fig3-Flow_chart}, the SHG effect of each magnetic structure can be classified, as shown in Fig. \ref{Fig4-MagnDat}. We {find} that 496 magnetic structures process the SHG effect, and 451 magnetic structures exhibit the LMO effect. {Out of the 496 magnetic structures with SHG effect, 100 of them} also have LMO effect. The 496 magnetic structures with SHG effect can be further {categorized} into the above seven types, as illustrated in Fig. \ref{Fig4-MagnDat}b. The clarification of the SHG and LMO effects for all magnetic structures in MAGNDATA is {compiled} into a database which is presented in Supplementary Note 6.

\begin{figure*} [!htbp]
\centering
\includegraphics[width=0.95\textwidth]{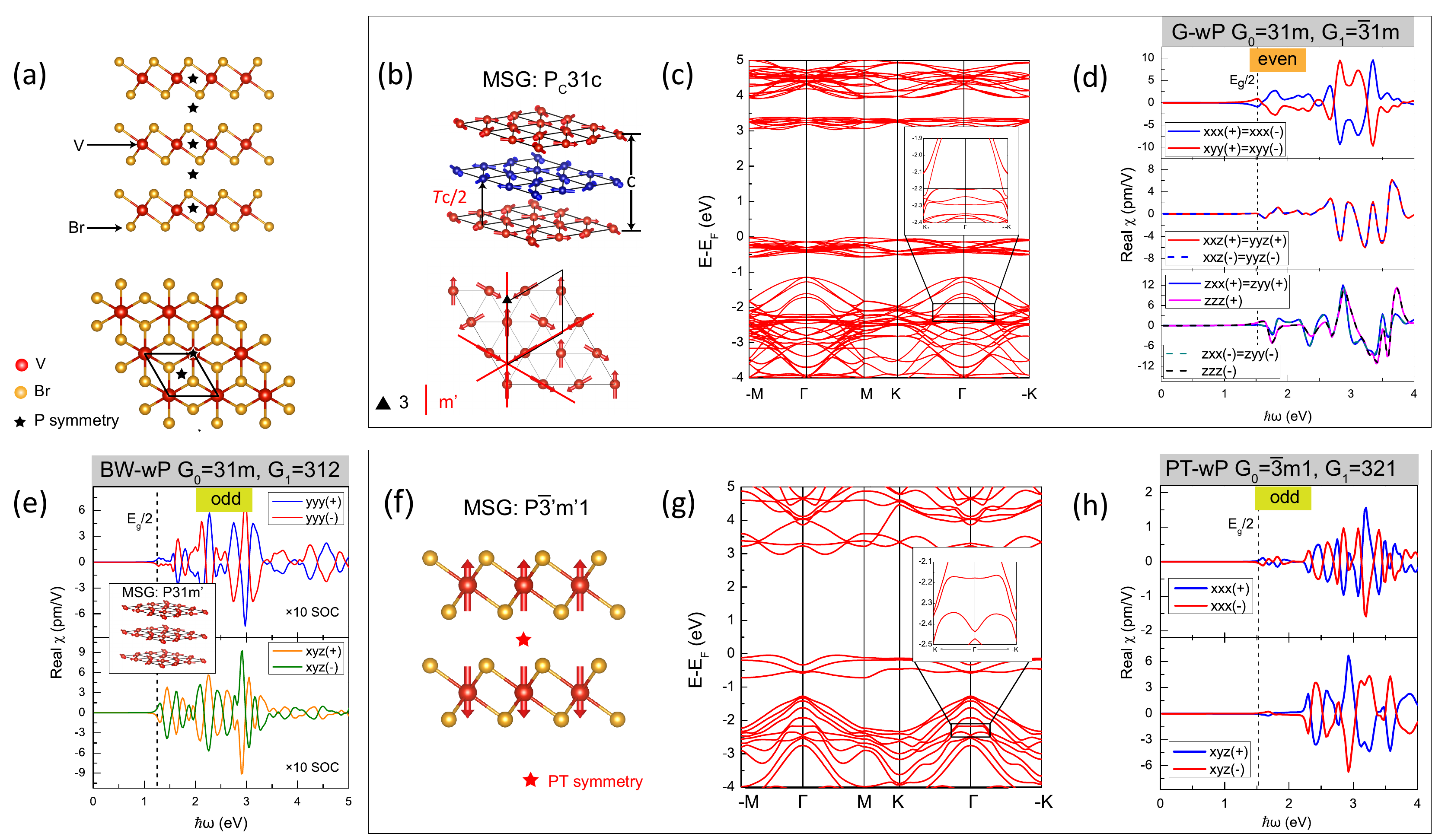} %
\caption{\textbf{Calculation results of VBr$_2$.} (\textbf{a}) {Crystal structure of parent phase of bulk} VBr$_2$ (upper panel: side view, lower panel: top view). (\textbf{b}) Magnetic structure (upper panel: side view, lower panel: top view), (\textbf{c}) band structures and (\textbf{d}) nonzero SHG susceptibilities of in-plane frustrated bulk VBr$_2$. (\textbf{e}) The odd SHG susceptibilities of bulk VBr$_2$ whose interlayer {magnetic moments are parallel aligned}. The SOC strength {is increased by an order of magnitude} to make the odd part of SHG susceptibilities more significant. (\textbf{f}) Magnetic structure (side view), (\textbf{g}) band structures and (\textbf{h}) nonzero SHG susceptibilities of bilayer VBr$_2$ with A-type AFM magnetic structure. The black rhombuses in (\textbf{a}) and (\textbf{b}) mean the unit cell and magnetic unit cell, respectively. The ``+" and ``$-$" signs in (\textbf{d}), (\textbf{e}) and (\textbf{h}) {denote} the positive and negative magnetic structures.}
\label{Fig-VBr2}
\end{figure*}

Guided by this classification, we {will discuss} two examples to {validate our isomorphic theory and classification method}. The first one {involves} VBr$_2$, which suggests that the magnetic structure can induce the {even-time-reversal} SHG tensors individually. In the second example, {we will study the SHG effect in} AFM materials RMnO$_3$ (R = Sc, Y, In, {Dy}, Ho, Er, Tm, Yb and Lu) with diverse magnetic structures.

\noindent \textbf{Example 1: Anomalous SHG in bulk VBr$_2$}

{It was previously presumed that SHG effect induced by magnetic structure would reverse upon switching the magnetic order due to the inherent oddness of magnetic order under time reversal symmetry.}  Here, we {present a typical example violating such assumption in bulk VBr$_2$ \cite{RN3344, RN3345, RN3340}}. As depicted in Fig. \ref{Fig-VBr2}a, the parent phase of bulk VBr$_2$ {processes the} inversion symmetry (SG: $P\bar{3}m1$), and one V atom has six nearest V atoms forming equilateral triangles.
Therefore, {in-plane} geometrical frustration is expected {accompanying with the inversion symmetry breaking when the magnetic interactions are antiferromagnetic}, as illustrated in Fig. \ref{Fig-VBr2}b. {Additionally}, the magnetic moments of the adjacent interlayer V atom are opposite {to each other}, resulting to the two adjacent layers be connected by $T\mathbf{c}/2$ symmetry.  {Its} MSG is $P_C31c$ ($P_C$ means the BW Bravais lattices), as presented in Fig. \ref{Fig-VBr2}b. According to the classification rule, the SHG type of bulk VBr$_2$ {is classified as} G-wP, which {is induced by the magnetic structure and it is even under time-reversal symmetry}.


The band structures of bulk VBr$_2$ are shown in Fig. \ref{Fig-VBr2}c. The bands split because of the breaking inversion symmetry and $PT$ symmetry, and the band energies of $+\mathbf{k}$ and $-\mathbf{k}$ points are equivalent due to the ``effective" time-reversal symmetry, {as shown in} the inset of Fig. \ref{Fig-VBr2}c. The nonzero even components of SHG tensor are constrained by $31m$ symmetry, {which means the independent elements are}  $\chi_{xxx}^{\text{even}}$ ($=-\chi_{xyy}^{\text{even}}=-\chi_{yxy}^{\text{even}}$), $\chi_{xxz}^{\text{even}}$ ($=\chi_{yyz}^{\text{even}}$), $\chi_{zxx}^{\text{even}}$ ($=\chi_{zyy}^{\text{even}}$) and $\chi_{zzz}^{\text{even}}$. The calculated SHG results are presented in Fig. \ref{Fig-VBr2}d, which {are} consistent with the above symmetry analysis. {Furthermore}, the SHG effect is even with time-reversal symmetry, {meaning that reversing} the magnetic order does not lead to the reversion of SHG susceptibilities [Fig. \ref{Fig-VBr2}d]. 

\begin{figure*}[!t]
\centering
\includegraphics[width=1.0\textwidth]{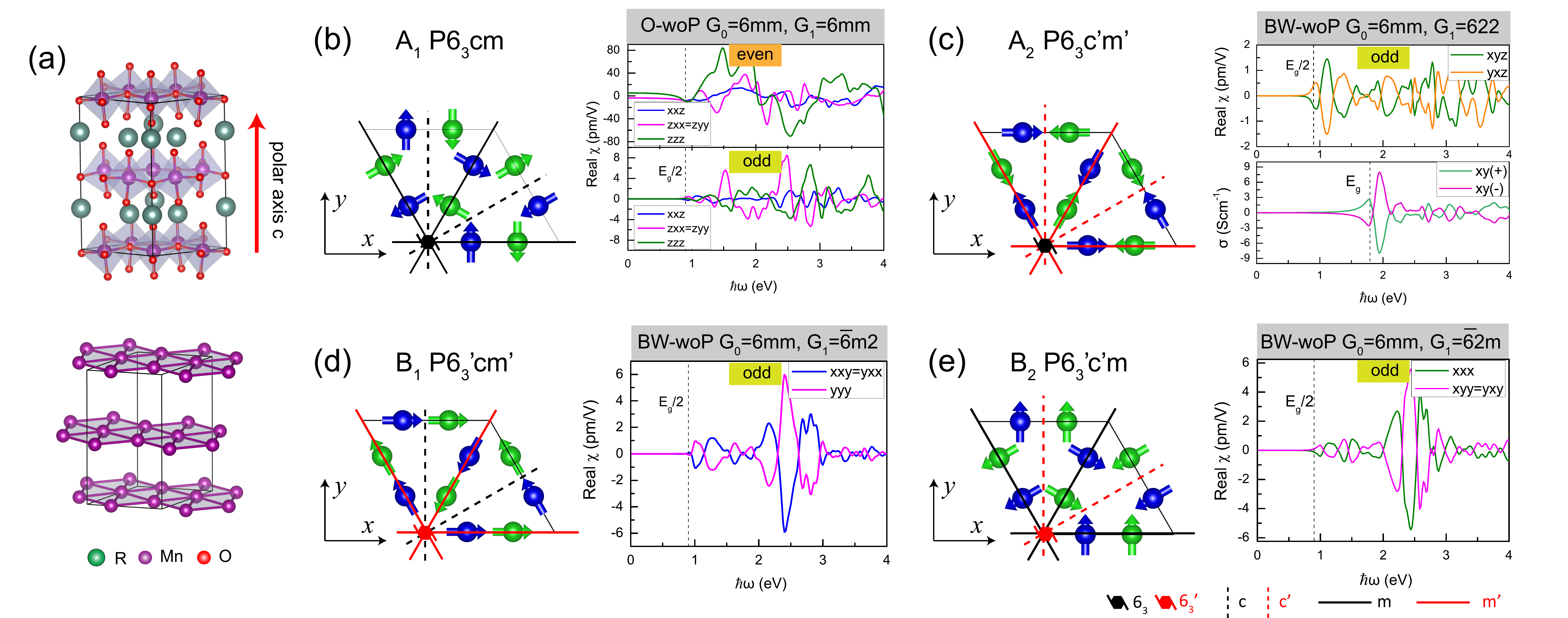}
\caption{\textbf{Magnetic structures and SHG properties of RMnO$_3$ (R = Sc, Y, In, Dy, Ho, Er, Tm, Yb and Lu) with different magnetic structures.} (\textbf{a}) Crystal structure of RMnO$_3$. {Magnetic structures of RMnO$_3$ with (\textbf{b}) $A_1$ ($P6_3cm$), (\textbf{c}) $A_2$ ($P6_3c'm'$), (\textbf{d}) $B_1$ ($P6_3'cm'$) and (\textbf{e}) $B_2$ ($P6_3'c'm$) phases. The blue and green balls in (\textbf{b})-(\textbf{e}) mean the Mn$^{3+}$ ions at different layers (top view). The right panels in (b)-(e) are the SHG susceptibilities of YMnO$_3$ with corresponding magnetic structures in the left. The lower panel of (\textbf{c}) is the calculated antisymmetric photoconductivity. The detailed magnetic and band structures are given in Supplementary Note 5.3.}}
\label{Fig7-YMnO3}
\end{figure*}

The G-wP type SHG {effect} originates from {the magnetic structures}, and it is even with time-reversal symmetry. {These two characteristics are the distinctive features of the G-wP type SHG effect}. The anomalous SHG effect in bulk VBr$_2$ refreshes the conventional understanding that magnetism can only contribute to the odd SHG tensors. In addition, the G-wP type SHG effect is not unusual in real materials, as 76 magnetic structures in the MAGADATA database {are proposed to exhibit G-wP SHG effect} (Fig. \ref{Fig4-MagnDat}b). Therefore, the classification method is a powerful tool to explore exotic SHG phenomena in {magnetically ordered materials}.

If the magnetic moments of V atoms in the adjacent interlayers are {parallel aligned}, as exhibited in the inset of Fig. \ref{Fig-VBr2}e, the MSG of this magnetic structure is $P31m'$. This magnetic structure leads to the BW-wP type SHG effect according to Fig. \ref{Fig3-Flow_chart}. The even and odd SHG tensors coexist, and both originate from magnetic structure. The even SHG susceptibilities are basically the same as those in Fig. \ref{Fig-VBr2}d, and the odd parts are shown in Fig. \ref{Fig-VBr2}e. We further investigate the SHG effect of bilayer VBr$_2$ with A-type AFM magnetism, {as} depicted in Fig. \ref{Fig-VBr2}f. Due to $PT$ symmetry, its bands are {doubly} degenerate, and the band energies of $+\mathbf{k}$ and $-\mathbf{k}$ points are not equivalent ({as illustrated in} the inset of Fig. \ref{Fig-VBr2}g). Since its MSG is $P{\bar{3}}'{m}'1$, {the corresponding} SHG effect belongs to the PT-wP type. Its SHG tensor has only the odd part, and the SHG tensor is constrained by the 32 ($D_3$) symmetry. The calculated SHG coefficients are shown in Fig. \ref{Fig-VBr2}h, which are consistent with our symmetry analysis. Indeed, the SHG effect {in bilayer VBr$_2$ with A-type AFM magnetism} reverses with the magnetic order, similar to that of bilayer CrI$_3$ \cite{RN3002, RN3003}.  

\noindent \textbf{Example 2: SHG effect of RMnO$_3$ with various magnetic structures}

The parent phase of RMnO$_3$ (R = Sc, Y, In, {Dy}, Ho, Er, Tm, Yb and Lu) usually adopt the non-centrosymmetric structure with SG $P6_3cm$, as presented in Fig. \ref{Fig7-YMnO3}a. The magnetism primarily arises from the Mn$^{3+}$, forming {approximately} equilateral triangles, as illustrated in the bottom panel of Fig. \ref{Fig7-YMnO3}b. Below the N\'eel temperature, the strong super-exchange leads to {120$^\circ$ arrangement of the spins of Mn$^{3+}$ in the basal plane, and small displacements of Mn$^{3+}$ ions (occupy $6c$ positions, see Supplementary Table VIII) break the triangular frustration}.

According to the MAGNDATA database and Refs. \cite{RN3336, RN3337, RN3605, RN3322, RN3601, RN3336, RN3603, RN3604}, RMnO$_3$ can exhibit various magnetic structures, such as $A_1$, $A_2$, $B_1$ and $B_2$ phase presented in Fig. \ref{Fig7-YMnO3}b-e. For a specific hexagonal manganite, the ground magnetic order always belongs to one of these states, but can be manipulated among them {under certain conditions}.
{The pioneering works by Fiebig et al. \cite{RN3605, RN3322, RN3601, RN3336, RN3603, RN3604} have proven the SHG effect is a powerful tool for investigating the magnetic structures of RMnO$_3$. In this section, we aim to further validate our theory and method with RMnO$_3$. This existing foundations enabled our current advances, and is instrumental in verifying the validity of our isomorphic theory and classification method.}

The inversion symmetry breaking of the parent phases leads {all RMnO$_3$} to exhibit SHG effects. The SHG effect of the $A_1$ phase belongs to the O-woP type, and the SHG effects of other three phases {falls into} the BW-woP type. The even SHG tensors of the four magnetic phases possess the same {characteristics}, which are constrained by PG $G_0=6mm$. {However}, the odd SHG tensors differ for every magnetic phase,  {due to the different symmetries constraints as shown in Fig. \ref{Fig7-YMnO3}b-e}. The calculated nonzero SHG {susceptibilities} of YMnO$_3$ with the four magnetic structures are depicted in Fig. \ref{Fig7-YMnO3}b-e, which are consistent with the symmetry analysis results (see Supplementary Table X) {and previous studies \cite{RN2843,RN3605, RN3322, RN3601, RN3336, RN3603, RN3604, RN3337}}. The magnitude of the even SHG susceptibilities are almost one order {of magnitude} larger than those of odd parts {in RMnO$_3$}, as presented in Fig. \ref{Fig7-YMnO3}b, which may be caused by the weak spin-orbit coupling (SOC) effect \cite{RN2842, RN3003, RN3300} in YMnO$_3$.

The in-plane even SHG components $\chi_{abc}^{\text{even}} (a,b,c\in \{x,y\})$ vanish {in} all magnetic phases {due to the $C_{2z}$ symmetry in $G_0=6mm$}. However, the isomorphic groups $G_1$ of the $B_1$ and $B_2$ phases break the {$C_{2z}$} symmetry, leading to the nonzero in-plane odd SHG components. While the in-plane SHG components of the $B_1$ and $B_2$ phases are different because these are {constrained} by distinct symmetries ($G_1(B_1)=\bar{6}m2$ and $G_1(B_2)=\bar{6}2m$). Besides, the symmetry of the $A_2$ phase (MPG: $6m'm'$) {permits} the LMO effect, with the calculated anti-symmetry photoconductivity {of YMnO$_3$} presented in Fig. \ref{Fig7-YMnO3}c. Despite {sharing} the same crystal structure, the different magnetic structures lead to distinctive SHG and LMO features, as summarized in Supplementary Table X. Therefore, these fingerprints can be used to distinguish the magnetic structures and domains of RMnO$_3$ via the SHG \cite{RN3327, RN3322, RN3325, RN3351, RN3605, RN3601, RN3336, RN3603, RN3604} and LMO effects.


\noindent \textbf{DISCUSSION}

In {magnetically ordered materials}, the SHG susceptibility can be expressed as a Tailor series expression {in terms of} the {magnetic order parameter} $\mathbf{M}$ \cite{RN3337} (such as magnetic moment in a ferromagnet, the N\'{e}el vector in a collinear antiferromagnet and the chirality in a non-collinear antiferromagnet):
\begin{equation}
\chi_{ijk}^{(2)} \text{=}\chi_{ijk}^{(2)} ({{\mathbf{M}}^{0}})+\chi_{ijk}^{(2)} ({{\mathbf{M}}^{1}})+\chi_{ijk}^{(2)} ({{\mathbf{M}}^{2}})+\cdots,
\end{equation}
where $\chi_{ijk}^{(2)}({{\mathbf{M}}^{0}})$ is independent on magnetic order, corresponding to the SHG effect only contributed by crystallographic asymmetry. The odd{-order} Tailor expansion terms, {including} $\chi_{ijk}^{(2)}({{\mathbf{M}}^{1}})$, $\chi_{ijk}^{(2)}({{\mathbf{M}}^{3}})$, $\chi_{ijk}^{(2)} ({{\mathbf{M}}^{5}})$, $\cdots$ contribute to the odd SHG tensors ($c$-type). {As stated in Ref.} \cite{RN2843}, the even-type SHG effect ($i$-type) refer to non-invariance under spin reversal, {allowing the even-order magnetic coupling} ($\chi_{ijk}^{(2)} ({{\mathbf{M}}^{2}})$, $\chi_{ijk}^{(2)}({{\mathbf{M}}^{4}})$, $\chi_{ijk}^{(2)} ({{\mathbf{M}}^{6}})$, $\cdots$) to contribute SHG effect. In the PT-wP type SHG effect, such as bilayer CrI$_3$ \cite{RN3002, RN3003}, the lowest Tailor term is $\chi ({{\mathbf{M}}^{1}})$, and only the odd-order terms exist. While for the G-wP type SHG effect, such as bulk VBr$_2$, the lowest Tailor term is $\chi ({{\mathbf{M}}^{2}})$, however the zero-order $\chi ({{\mathbf{M}}^{0}})$ and the linear-order term $\chi ({{\mathbf{M}}^{1}})$ are forbidden by symmetry.


{It is noted that our study is mainly focused on the SHG effect induced by the bulk electric dipole. The classification method can also be extended to the SHG effect arising from surface electric dipole provided that the surface crystal and magnetic structures are given.}
In centrosymmetric materials, the magnetic-dipole-type and electric-quadrupole-type SHG are permitted {by symmetry}. For example, the SHG effect in Cr$_2$O$_3$ \cite{RN2498} and monolayer CrSBr \cite{RN3196}.
{Even thorough, there are prior works on magnetic-dipole SHG effect by Hanamura and Valent\'i \emph{et al.} \cite{RN2849, RN2850, RN2496, RN2864, RN2499} using semi-classical methods to treat the magnetic-dipole moment operator,} evaluating the magnetic-dipole-type and electric-quadrupole-type SHG effect in first-principles calculations {remain tough issues. Because it is very complicated to
accurately treat the magnetic-dipole and electric-quadrupole operators, and capture their interplay with electric fields in density functional theory (DFT)}.
Hence, the magnetic-dipole-type and electric-quadrupole-type SHG effects are {not numerically studied} in this work.
{However}, we think our isomorphic group method can also be applied to investigate the characteristics of these two SHG effects, {which needs to be further studied}.

According to the classification {presented in} Table \ref{Tab3-ClassificationSHG}, the physical {mechanisms} of SHG effect {can be classified into three main aspects: the even SHG effect arising from crystal asymmetry, even SHG effect caused by magnetic structure and odd SHG effect caused by magnetic structure}. As summarized in Table \ref{Tab5-SHGdiffer}, we will compare them in the following. Specifically, the magnetic structure can contribute to both the odd and even SHG tensors, which vanish when the {magnetic order} or SOC effect in the coplanar {magnetically ordered materials} disappear. {In this case}, only the even SHG tensors contributed by the crystal structure asymmetry survive. For semiconducting {magnetically ordered materials}, ${{\chi}^{\text{even}}}$ is nonzero, while ${{\chi}^{\text{odd}}}$ disappears in the zero frequency limit ($\omega \to 0$, see Supplementary Note 5.5 for detail).
The conditions to obverse them individually are in non{magnetically ordered materials}, G-wP type and PT-wP type {magnetically ordered materials}, respectively, as listed in Table \ref{Tab5-SHGdiffer}.
The three parts in Table \ref{Tab5-SHGdiffer} can coexist in {magnetically ordered materials} whose parent phases have no inversion symmetry, corresponding to BW-woP and O-woP types SHG effect.
{It should be noted} that the seven types of SHG identified in our work are required further experimental verification. Therefore, we encourage researchers to perform experiments and simulations to validate our {proposed} classification and {uncover distinctive} SHG characteristics in {magnetically ordered materials}.


\begin{table}[!htpb]
\centering
\caption{\textbf{Distinctive features of different physical {mechanisms of SHG effect} in {magnetically ordered materials}.}}
\label{Tab5-SHGdiffer}
\begin{threeparttable}
\begin{tabular} {m{2.0cm}|m{2.0cm}<{\centering}|m{2cm}<{\centering}|m{1.8cm}<{\centering}} 
  \hline
  \hline
	&	Even from crystal asymmetry	&	Even from {magnetic structure}	&	Odd	from {magnetic structure} \\ \hline
Without magnetic order	&	$\checkmark$	&	$\times$	&	$\times$	\\ \hline
Without SOC effect$^{(1)}$	&	$\checkmark$	&	$\times$	&	$\times$	\\ \hline
Zero frequency limit$^{(2)}$	&	$\checkmark$	&	$\checkmark$	&	$\times$	\\ \hline
Reverse {magnetic order}	&	Invariant	&	Invariant	&	Reverse	\\ \hline
{Individually existing condition} $^{(3)}$	&	Non{magnetically ordered materials}	&	G-wP	&	PT-wP	\\ \hline
\hline
\end{tabular}
  \begin{tablenotes}
    \footnotesize
    \item[(1)] For the coplanar {magnetically ordered materials}, see Supplementary Note 5.4 for more information.
    \item[(2)] For semiconducting {magnetically ordered materials}, see Supplementary Note 5.5 for more information.
    \item[(3)] {``Individually existing condition" refers to the symmetry requirements to observe the SHG effect arising from crystal asymmetry, even SHG effect caused by magnetic structure and odd SHG effect caused by magnetic structure, individually.}
   \end{tablenotes}
\end{threeparttable}
\end{table}

In conclusion, the classification of SHG effect in magnetically ordered materials is performed to reveal the symmetries and physical mechanisms. Specifically, the isomorphic group method is utilized to study the SHG tensor {for} every MPG. {This method simplifies} the SHG tensor characteristics under MPGs into nonmagnetic PGs, {revealing the symmetry rules governing SHG effect}. {Furthermore}, a tensor dictionary containing SHG and LMO effects is created, {serving as a guidance for} experimentally probing magnetic structures. Subsequently, SHG effect is classified into seven {distinct} types based on the symmetries of the magnetic and parent {crystal structures}. Assisted by this classification, {a database cataloging SHG and LMO effects of materials in} MANGDATA is established. The proposed classification method {builds a close relationship} between the SHG effect and magnetic structures, and {enhance} the understanding of the SHG effect in {magnetically ordered materials}.
\textbf{}

\textbf{}
\footnotesize{
{\setlength{\parindent}{0cm}

\textbf{METHODS}

\textbf{First-principles calculations}

The first-principles calculations based on DFT are performed using the VASP software package \cite{RN1434, RN1433}. Moreover, the general gradient approximation (GGA) according to the Perdew-Burke-Ernzerhof (PBE) functional is employed, and the spin-orbit coupling effects are considered for all the materials. $U_{eff}$= 3.0 eV (5.0 eV) is set for V (Mn) atoms in VBr$_2$ (YMnO$_3$) calculations. The Bloch wave functions are iteratively transformed into maximally localized Wannier functions using the Wannier90 package \cite{RN149, RN772}. The SHG susceptibilities and antisymmetric photoconductivities are calculated using our program WNLOP (Wannier Non-Linear Optics Package) based on the effective tight-binding (TB) Hamiltonian obtained by Wannier90.

\textbf{Symmetries analysis of {magnetically ordered materials}}

The symmetries of {magnetically ordered materials} are determined by the \href{https://stokes.byu.edu/iso/findsym.php}{FINDSYM} website \cite{RN3332}, the and MPGs and MSGs are analyzed with {the help of} \href{https://www.cryst.ehu.es/#msgtop}{Bilbao Crystallographic Server} \cite{RN3352}. The magnetic structures are visualized using VESTA software \cite{RN3349}.

\textbf{}

\textbf{}
\textbf{DATA AVAILABILITY}

All data required to support the conclusions are presented in the paper. Additional data related to this paper can be requested from the authors.

}
\bibliographystyle{apsrev4-1}
\bibliography{Reflatex}

\footnotesize{
{\setlength{\parindent}{0cm}
\textbf{Acknowledgments}

The authors acknowledge the High-performance Computing Platform of Anhui University for providing computing resources. We thank Haowei Chen (Tsinghua University), Shu-Hui Zhang (BUCT), Bo-Lin Li (HFIPS), Yang Gao (USTC), Yang Gao (AUST), Chong Wang (CMU), Jian Zhou (Xi'an Jiaotong University), Jiang Zeng (Hunan University), Kai Huang (UNL),  {Weikang Wu (Shandong University) and Y. J. Jin (Nanyang Technological University)} for their useful discussions. This work is supported by the National Key R\&D Program of China (Grants No. 2022YFA1403700), the National Natural Science Foundation of China under No. 12204009, 12204003, 11947212, 11904001, Natural Science Foundation of Anhui Province under No. 2208085QA08 and 2008085QA29, in part by the Joint Funds of the National Natural Science Foundation of China and the Chinese Academy of Sciences Large-Scale Scientific Facility under Grant No. U1932156. R.-C.X. acknowledge the startup foundation from Anhui University.

\textbf{Author contributions}

R.-C.X. conceived the project with H.L. and H. J.. R.-C.X. edited the code and performed the symmetry analysis and first-principles calculations. R.-C.X., H.J., and H.L. discussed the results and the writing. The manuscript was written through the contributions of all authors. All authors have approved the final version of the manuscript.

\textbf{Competing interests}

The authors declare that they have no competing financial or non-financial interests.

\textbf{Additional information}

\textbf{Supplementary information} The online version contains supplementary material available at https://doi.org/xxxx.

}}

\end{document}